\documentclass[pre,a4paper,twocolumn]{revtex4}
\usepackage[T1]{fontenc}
\usepackage[cp1250]{inputenc}
\usepackage{graphicx}
\usepackage{amsmath}
\usepackage{psfrag}

\newcommand{\bq}{\begin{equation}}
\newcommand{\eq}{\end{equation}}
\newcommand{\ba}{\begin{eqnarray}}
\newcommand{\ea}{\end{eqnarray}}

\begin{document}
\title{\bf Approaching the thermodynamic limit in equilibrated scale-free networks}
\author{B. Waclaw$^1$}
\author{L. Bogacz$^2$}
\author{W. Janke$^{1,3}$}

\affiliation{$^1$Institut f\"ur Theoretische Physik, Universit\"at Leipzig,
Postfach 100\,920, 04009 Leipzig, Germany \\
$^2$Department of Information Technologies, Faculty of Physics, Astronomy and Applied Informatics, Jagellonian University, Reymonta 4, 30-059 Krak\'ow, Poland \\
$^3$Centre for Theoretical Sciences (NTZ), Universit\"at Leipzig, Germany
}

\begin{abstract}
We discuss how various models of scale-free complex networks approach their limiting properties when the size $N$ of the network grows. We focus mainly on equilibrated networks and their finite-size degree distributions. Our results show that the position of the cutoff in the degree distribution, $k_{\rm cutoff}$, scales with $N$ in a different way than predicted for $N\to\infty$, that is subleading corrections to the scaling $k_{\rm cutoff}\sim N^{\alpha}$ are strong even for networks of order $N\sim 10^9$ nodes. We observe also a logarithmic correction to the scaling for degenerated graphs with the degree distribution $\pi(k)\sim k^{-3}$. On the other hand, the distribution of the maximal degree $k_{\rm max}$ may have a different scaling than the cutoff and, moreover, it approaches the thermodynamic limit much faster. We argue that $k_{\rm max}\sim N^{\alpha'}$ with the exponent $\alpha'=\min(\alpha,1/(\gamma-1))$, where $\gamma$ is the exponent in the power law $\pi(k)\sim k^{-\gamma}$. We present also some results on the cutoff function and the distribution of the maximal degree in equilibrated networks.
\end{abstract}

\maketitle

\section{Introduction}
Recent progress in understanding the structure and function of complex networks \cite{ref:cn} has been largely influenced by the application of statistical methods of modern physics. The statistical mechanics of networks \cite{ref:snd-stat-mech,ref:mejn,ref:homnasz,ref:bianconi,ref:bernard,ref:dr2,ref:kr}, if restricted to structural properties, deals with two classes of problems.
In the first one, one considers networks being in a sort of equilibrium \cite{ref:snd-stat-mech,ref:homnasz,ref:bianconi,ref:bernard} where the concept of statistical ensembles applies in a natural way. On the opposite, growing networks are usually treated by rate-equation formalism \cite{ref:dr2,ref:kr}. Both approaches, however, share one common thing: the majority of considered models is analytically solvable only in the thermodynamic limit. This causes several difficulties. First of all, a question arises about the statistical equivalence of different ensembles of networks like canonical or grand-canonical ensemble \cite{ref:snd-stat-mech}. Second, considering dynamical processes on networks, one can ask how they are influenced by approaching the thermodynamic limit in different ways \cite{castellano}. 

The third question, which we want to address in this paper, is how fast is the convergence towards the limiting values of some network features, for different models. This question is much more important than the analogous question in more traditional physics dealing with atoms or molecules, where the number of degrees of freedom is typically of order $10^{23}$. Here, on the contrary, the largest observed system is the World-Wide Web with $N \sim 10^9 - 10^{10}$ nodes. Many real networks are much smaller than this, typically having only $10^3 -10^4$ nodes.
Therefore, finite-size corrections to solutions obtained in the thermodynamic limit are much stronger and in many situations cannot be neglected.

As a most prominent example we will consider the degree distribution $\pi(k)$ which is of most importance for characterizing the network as well as many dynamical processes taking place on it. 
Finite-size effects are especially strong for scale-free networks, for which $\pi(k)\sim k^{-\gamma}$ exhibits a power law. Examples are the Internet, the WWW and many social and biological networks.
Since the power law cannot extend to infinity for finite size $N$, the degree distribution $\pi(k)$ must decrease rapidly above some characteristic value $k=k_{\rm cutoff}$. The cutoff scales typically as $k_{\rm cutoff}\sim N^\alpha$, where the exponent $\alpha$ depends on the network type. For $2<\gamma\leq 3$ when the second moment is divergent in the limit of large $N$, the cutoff appears either directly, or via moments of $\pi(k)$ in many applications, for instance in percolation or infection spreading \cite{ref:d1}, transport models \cite{ref:d2}, synchronization processes and others, see Ref.~\cite{ref:crit-phenom}. Because qualitative differences may show up for various values of $\alpha$, it is important to know the proper scaling of $k_{\rm cutoff}$ with $N$.

For a growing network like the Barab\'{a}si-Albert (BA) model \cite{ref:bamodel} or its generalization, the growing network with re-direction (GNR) model \cite{ref:kr}, the scaling is well established \cite{ref:kr2, ref:BWS}: for any $\gamma>2$, the exponent $\alpha=1/(\gamma-1)$. Moreover, it can be shown that for $N$ large enough $\pi(k)$ can be approximated as $\pi_{\infty}(k) w(k/N^\alpha)$, where $\pi_\infty(k)$ is the degree distribution for $N\to\infty$ and $w(x)$ is some function independent of $N$. This approximation holds even for quite small networks of order $10^3$ nodes because the convergence towards the limiting distribution is fast.

The situation is not so clear for equilibrated networks, that is networks in which evolution is governed by rewiring of existing connections rather than by adding new nodes, and thus can be regarded as being in a sort of equilibrium.
Although the scaling exponent $\alpha$ has been estimated for some models \cite{ref:dr3,ref:zbak2,ref:extr2,ref:bck}, the cutoff function $w(x)$ has not been determined so rigorously as for growing networks. 

In this paper we will show that the predicted scaling is far from being true even for quite large networks. The paper is organized as follows. In Sec. II we discuss three models of equilibrated networks whose large-$N$ behavior we want to study.
Section III starts with basic concepts of how to calculate finite-size degree distributions and how to extract the behavior of the cutoff function. Then we consider two models, for which precise, semi-analytical results are available for sizes up to $N\sim 10^9$.
In Sec. IV we discuss a more complicated model and show how to simulate it on a computer. Section V is devoted to a relation between the cutoff and the maximal degree. The paper is closed with a short summary in Sec. VI.

\section{Models and their properties in the thermodynamic limit}
\label{sec:models}

We shall start from defining three different models of equilibrated networks, whose finite-size properties will be further examined. 
The word ``equilibrated'' means that networks (graphs in mathematical language) are maximally random under given constraints. These constraints are what defines the statistical ensemble of networks. The statistical ensemble consists of a set of states -- graphs $\{g_i\}$, and a set of their statistical weights $\{W(g_i)\}$. This means that every graph $g$ has a probability of occurrence proportional to $W(g)$.
Every physical quantity $X$ is then defined to be the average over the set of graphs: $\left<X\right> = \sum_g W(g) X(g)/\sum_g W(g)$.
Changing the set of graphs and/or the weights, one can obtain different models of random graphs.
Three such ensembles will be considered in this paper and are defined below. In all cases, we assume that the graphs are undirected, have labeled nodes, fixed number of nodes $N$ and links $L$, and that the weight $W(g)$ of every labeled graph has a product form:
\bq
	W(g) = \prod_{i=1}^N p(k_i), \label{wprod}
\eq
where $k_i$ is the degree of the $i$th node of graph $g$ and $p(k)$ is some arbitrary node-weight function. 
The product measure (\ref{wprod}) turns out to be very convenient to obtain the desired degree distributions in the thermodynamic limit by tuning the function $p(k)$.
In what follows, we will assume that all models have the same distribution $\pi(k)$ in the thermodynamic limit.
Different models will be specified by restricting the set of possible graph shapes and/or $p(k)$.

{\bf Equilibrated simple graphs}. We consider graphs without self- and multiple connections. 
Denoting by $\pi_\infty(k)$ the degree distribution for $N\to\infty$, we have:
\bq
	\pi_\infty(k) = \frac{p(k)}{k!} e^{A+Bk}, \label{pik}
\eq
as follows from Refs.~\cite{ref:snd-stat-mech,ref:bjk1}. The constants $A,B$ are chosen to have $\sum_k \pi_\infty(k)=1$ and $\sum_k k \pi_\infty(k) = 2L/N = \bar{k}$. Equation (\ref{pik}) can be used in two different ways. First, it tells us what $\pi_\infty(k)$ would be for  given $p(k)$ and the average degree $\bar{k}$. Second, it allows us to calculate the weight $p(k)$ which will produce the desired degree distribution $\pi_{\rm des}(k)$ in the thermodynamic limit. In the latter case, if $N,L$ are chosen so that the average degree $\bar{k}$ is equal to $\left<k\right> = \sum_k k \pi_{\rm des}(k)$, then $B=0$ and the choice $p(k)=\pi_{\rm des}(k) k!$ implies that $\pi_\infty(k)=\pi_{\rm des}(k)$. 

{\bf Equilibrated multigraphs}. We again assume the product weight (\ref{wprod}), but now we accept also degenerated graphs, that is graphs with multiple- and self-connections. It can be shown \cite{myphd,ref:zbak2,ref:snd-stat-mech} that the partition function of the system, being the sum over all configurations, assumes the form:
\bq
	Z(N,L) = 
	\sum_{k_1=0}^\infty \cdots \sum_{k_N=0}^\infty \frac{p(k_1)}{k_1!} \cdots \frac{p(k_N)}{k_N!} \delta_{2L,k_1+\dots+k_N}, \label{Zpseudo}
\eq
so it is equivalent to that of the balls-in-boxes model \cite{ref:bib} or the zero-range-process model \cite{ref:zrp} with weights $p(k)/k!$. The same formula (\ref{pik}) as for simple graphs holds for the degree distribution in the limit $N\to\infty$. 

{\bf Equilibrated trees}. This ensemble consists of all labeled, connected tree graphs with $N$ nodes. 
In addition, we assume that one node is distinguished by a ``stem'' attached to it, which is convenient from a mathematical point of view, but it does not change the large-$N$ behavior.
Such ``planted'' trees can be treated in a special way and a number of quantities can be calculated analytically \cite{ref:bck, ref:bbjk}. For instance, 
in the thermodynamic limit the degree distribution is given by
\bq
	\pi_\infty(k) = \frac{p(k)}{(k-1)!} e^{A+Bk}. \label{piktree}
\eq
Note that we have $(k-1)!$ in the denominator, in contrast to Eq.~(\ref{pik}). 
Now, the desired degree distribution $\pi_{\rm des}(k)$ can be obtained only if $\sum_k k \pi_{\rm des}(k)=2$, because the average degree
for trees approaches $2$ for $N\to\infty$. When this criterion is fulfilled, then assuming $p(k)=\pi_{\rm des}(k)(k-1)!$ one obtains $\pi_\infty(k)=\pi_{\rm des}(k)$.

All these models share one common property: 
the degree distribution $\pi_\infty(k)$ in the thermodynamic limit is proportional to $p(k)$.
Therefore, by choosing $p(k)\sim k! k^{-\gamma}$ (for graphs) or $\sim (k-1)! k^{-\gamma}$ (for trees),
one can make these networks scale-free.
In this paper we shall stick to the following choice for the degree distribution in all models:
\bq
	\pi_\infty(k) = \frac{(\gamma-1)\Gamma(2\gamma-3)}{\Gamma(\gamma-2)} \frac{\Gamma(k+\gamma-3)}{\Gamma(k+2\gamma-3)} \sim k^{-\gamma},	\label{piDMS}
\eq
for $k>0$, and $\pi_\infty(0)=0$. 
This is precisely the degree distribution $\pi_\infty(k)$ in the GNR model of a growing tree mentioned above.
The average degree for this distribution is $\left<k\right>=2$, therefore the equilibrated trees will approach $\pi_\infty(k)$ in a natural way if $p(k)=\pi_\infty(k)(k-1)!$.
As to equilibrated simple graphs and multigraphs, one then has to ensure that $p(k)=\pi_\infty(k)k!$ and $\bar{k}=2L/N\to 2$ for $N\to\infty$, which can be simply done by assuming $L=N$.
The purpose of choosing this particular distribution is that since finite-size effects in the GNR model are known \cite{ref:BWS}, we can compare what happens if networks are equilibrated but have the same $\pi_\infty(k)$ as growing ones.

\section{Degree distribution for a finite network}
\label{sec:degr}

In the previous section we discussed the behavior of the three models in the thermodynamic limit. Now we shall ask, how the degree distribution looks like for $N<\infty$.
Assume that $\pi_N(k)$ and $\pi_\infty(k)$ are degree distributions for finite $N$ and $N\to\infty$,
respectively.
It is convenient to write the finite-size distribution $\pi_N(k)$ as a product of $\pi_\infty(k)$ being $N$-independent, and some cutoff function $w(N,k)$ depending explicitly on the size $N$:
\bq
	\pi_N(k) = \pi_\infty(k) w(N,k).
\eq
The function $w(N,k)$ is model dependent. In Ref.~\cite{ref:BWS} it has been found that for the GNR model,
moments of $w(N,k)$ scale as follows:
\bq
	\mu_m \equiv \sum_k w(N,k) k^m \propto N^{\alpha(m+1)} (1+\mathcal{O}(N^{-\alpha})) \label{kmw}
\eq
with the cutoff exponent $\alpha$ defined in the introduction and equal to $\alpha=1/(\gamma-1)$. This means that for sufficiently large $N$ the cutoff function depends effectively on a single rescaled variable $x=k/N^\alpha$:
\bq
	w(N,k) \cong w(k/N^\alpha).	\label{wxdef}
\eq
For the GNR model, it is possible to find an explicit form of $w(x)$ for some values of $\gamma$. It is usually complicated, but it always has the following large-$x$ behavior:
\bq
	\ln w(x) \sim -x^{1/(1-\alpha)} = -x^\eta, \label{wxasympt}
\eq
where we defined the exponent $\eta=1/(1-\alpha)$.
Formulas (\ref{kmw}), (\ref{wxdef}) and (\ref{wxasympt}) were verified numerically \cite{myphd} for $N$ of order $10^3-10^4$, and an excellent agreement has been found.
In this paper we ask, to what extent these relations are valid for equilibrated networks.
In particular, if the scaling (\ref{wxdef}) still holds, the moments $\mu_m$ will behave as  $\mu_m \sim N^{\alpha(m+1)}$ for very large networks. The convergence towards this asymptotic behavior can be, however, different from that of growing networks. 
Therefore, we can relax the constraint that the sub-leading term decays with the power $\alpha$ and generally
expect the following large-$N$ behavior:
\bq
	\mu_m(N) \cong a N^{\alpha(m+1)} (1+b N^{-\beta}), \label{expectedmu}
\eq
with some constants $a>0,b\neq 0$ and $\beta>0$, so that the subleading correction decays like $N^{-\beta}$ with $\beta$ not
necessarily equal to $\alpha$, as it was in Eq.~(\ref{kmw}).

\subsection{Multigraphs}
\begin{figure}
	\includegraphics[width=7.5cm]{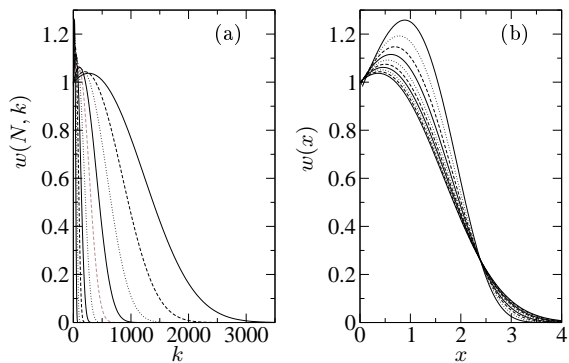}
	\caption{\label{fig0}(a) Plots of $w(N,k)$ for multigraphs of sizes between $100$ (the narrowest peak) and $51200$ (the widest peak). (b) The same data but plotted against the rescaled variable $x=k/(N\ln N)^{1/2}$. The highest peak shows data for $N=100$.}
\end{figure}

\begin{figure}
	\includegraphics*[width=8cm]{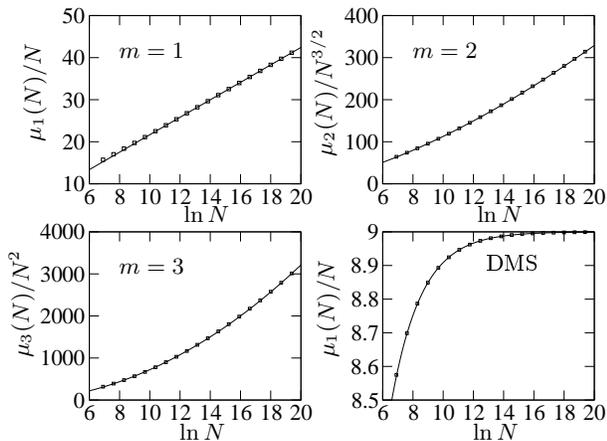}
	\caption{\label{fig1}Plots of $\mu_m(N)/N^{(m+1)/2}$ calculated from Eq.~(\ref{mm}) (squares), for degenerated graphs with $\gamma=3$ and for the GNR tree network with the same $\pi_\infty(k)$ for comparison. Solid lines display functions: (\ref{expectedmu}) for lower-right panel, and (\ref{logmu}) for other panels, with parameters fitted to data points.}
\end{figure}

We shall start from multigraphs. As we said, we assume that the average degree $\bar{k}$ is chosen to ensure $\sum_k k \pi_\infty(k)=\bar{k}$. This means that for a given number of nodes $N$, the number of links $L=L(N)$ is fixed.
In case of the distribution (\ref{piDMS}), $L=N$.
The partition function (\ref{Zpseudo}) becomes a function of $N$ only and can be rewritten as:
\bq
	Z(N) = \oint \frac{dz}{2\pi i} z^{-1-N\bar{k}} F^N(z), \label{Zpseudo2}
\eq
where the contour of integration encircles zero and $F(z)$ denotes the generating function for $\pi_\infty(k)$:
\bq
	F(z) = \sum_{k=0}^\infty \pi_\infty(k) z^k. \label{Fpseudo}
\eq
The degree distribution for finite $N$ can be calculated as follows \cite{ref:bbjk}:
\ba
	\pi_N(k) &=& \frac{p(k)}{NZ(N)} \frac{\partial Z(N)}{\partial p(k)} \nonumber \\
	& = &\pi_\infty(k) \frac{\oint \frac{dz}{2\pi i} z^{k-1-N\bar{k}} F^{N-1}(z)}{\oint \frac{dz}{2\pi i} z^{-1-N\bar{k}} F^N(z)} \nonumber \\
	&\equiv &\pi_\infty(k) w(N,k), \label{pikfinite}
\ea
where the cutoff function $w(N,k)$ is given by the above ratio of contour integrals. 
The integrals cannot be in general performed analytically for arbitrary $N$. 
However, the integration can be easily done numerically for different distributions $\pi_\infty(k)$ and sizes $N$ as follows. We choose the contour of integration to be $z=re^{i \phi}$, with $\phi\in (-\pi,\pi)$ and $r$ smaller than the radius of convergence of $F(z)$. Because the integrated function becomes concentrated around zero for large $N$, one does not need to integrate over the whole range of $\phi$. We have written a procedure in Mathematica, which finds the value of $r$, for which the integrated function has the broadest maximum at $\phi=0$.
This speeds up the convergence of numerical integration. Then, the procedure searches for $\phi_{\rm max}$ for which the function falls to a sufficiently small part (typically $10^{-10}$) of its maximal value. 
Then the function is integrated by means of the adaptive method over the range $(-\phi_{\rm max},\phi_{\rm max})$.
Let us consider first the distribution (\ref{piDMS}) for $\gamma=3$, when it reduces to
\bq
	\pi_\infty(k) = \frac{4}{k(k+1)(k+2)}, \label{piBA}
\eq
for which
\bq
	F(z) = \frac{z(3z-2)-2(z-1)^2 \ln (1-z)}{z^2}, \label{fz}
\eq
so that the radius of convergence is one.
Using the above numerical procedure we have calculated $w(N,k)$ for $N=100,200,\dots,51200$, shown in Fig.~\ref{fig0}a.
The theoretical value of $\alpha$, predicted for this model, should be $1/2$ \cite{ref:zbak2}, the same value comes from the correspondence to the zero-range process \cite{ref:evans2}. Therefore, for large $N$, plots of $w(N,k)$ for different sizes should collapse into a single curve in the rescaled variable $x=k/N^\alpha=k/N^{1/2}$. 
To check how fast the scaling is approached when $N$ grows, instead of using the cutoff function directly it is better to consider its moments $\mu_m$ which can be calculated as follows:
\bq
	\mu_m(N) = \frac{\oint \frac{dz}{2\pi i} z^{-1-N\bar{k}} F^{N-1}(z) f_m(z)}{\oint \frac{dz}{2\pi i} z^{-1-N\bar{k}} F^N(z)},
	\label{mm}
\eq
where the auxiliary function $f_m(z)$ is defined as
\bq
	f_m(z) = \left(z\frac{d}{dz}\right)^m \frac{1}{1-z}.
\eq
We evaluated numerically the integrals in Eq.~(\ref{mm}) using the same procedure as for the cutoff function, 
for $m=1,2,3$ and sizes $N=(1,2,\dots,2^{19})\times 1000$. The last value of $N$ is $\approx 5\times 10^8$. 
In Fig.~\ref{fig1} we show plots of $\mu_m(N)/N^{(m+1)/2}$, where the sub-leading behavior has been exposed by dividing the data by the asymptotic form of $\mu_m \sim N^{\alpha(m+1)}$. 
Since $N$ spans several orders of magnitude, the horizontal axis has a logarithmic scale  in order to make all points visible.
The first moment $\mu_1(N)/N$ for the GNR model of growing tree network, calculated as in Ref.~\cite{ref:BWS}, is also plotted for comparison.
It approaches its maximal value as $\sim N^{-0.5}$, which agrees with Eq.~(\ref{kmw}). 
But the moments for the degenerated graph show quite distinct behavior, namely they grow with $\ln N$ linearly or faster. 
This means that the subleading, power-like term $N^{-\beta}$ from Eq.~(\ref{expectedmu}) might turn into a leading behavior which takes place for $\beta\to 0$ and leads to a multiplicative, logarithmic correction.
To check this, we fitted the formula
\bq
	\mu_m(N)/N^{(m+1)/2} = A_m(\ln N)^{C_m} +B_m \label{logmu}
\eq
to data points in Fig.~\ref{fig1}. The fits are represented there as solid lines.
The ratio of obtained values $C_1:C_2:C_3 = 1.07:1.54:2.08 \approx 1:1.5:2$ suggests that the moments behave as
\bq
	\mu_m(N) \sim (N \ln N)^{(m+1)/2}.
\eq
In other words, the cutoff scales as $\sim (N\ln N)^{1/2}$. This is also confirmed in Fig.~\ref{fig0}b where the cutoff function is plotted in the rescaled variable $x=k/(N\ln N)^{1/2}$.
According to our knowledge, this logarithmic correction has not been observed before. 
Its origin cannot lie in the distribution $\pi_\infty(k)$ only, because the growing networks with the same
$\pi_\infty(k)$ do not have it, but it is the property of equilibrated graphs. Indeed, one can predict this scaling analytically, studying
the cutoff function $w(N,k)$:
\bq
	w(N,k) \cong W(N,k)/W(N,0),
\eq
where
\bq
	W(N,k) = \oint \frac{dz}{2\pi i} z^{k-1-N\bar{k}} F^{N}(z),
\eq
with $\bar{k}=2$ and $F(z)$ given by Eq.~(\ref{fz}). Following the lines of Sec.~6 from Ref.~\cite{ref:evans2} one can argue that the function under the integral is localized around $z=1$ and can be expanded at this point. Choosing the contour of integration $z=e^{i\phi}$ we obtain
\bq
	W(N,k) \cong \int_{-\epsilon}^\epsilon \frac{d\phi}{2\pi} e^{N\left[\frac{ik\phi}{N}+\phi^2(3-i\pi+\ln\phi^2)\right]}
\eq
with $\epsilon$ small enough. Inserting now $k=x(N\ln N)^{1/2}$ and $\phi=t(N\ln N)^{-1/2}$ we have
\bq
	W(N,x(N\ln N)^{1/2}) \cong (N\ln N)^{-1/2} \int \frac{dt}{2\pi} e^{itx - t^2 + \dots }
\eq
where higher terms are of order $\ln\ln N/\ln N$ and can be neglected for large $N$. Then the above formula becomes a Gaussian integral
and hence the cutoff function reads
\bq
	w(N,k)\approx e^{-\frac{1}{4}\left(\frac{k}{(N\ln N)^{1/2}}\right)^2},
\eq
with the predicted scaling.

Let us now go to $\gamma>3$. In this region, the exponent $\alpha$ should be again $1/2$ \cite{ref:evans2}. We calculated the first three moments $\mu_m(N)$ for $\gamma=3.5$ from Eq.~(\ref{mm}) for the same sizes $N$ as previously. Next, we fitted formula (\ref{expectedmu}) to the data points.
The best fit gives $\beta \approx 0.24$ for all three moments. This value is significantly greater than zero, thus the correction is now clearly of the power-like type. But the exponent $\beta \approx 1/4$ is small enough to produce large corrections even for moderate-size networks. For example, if $N=10^4$, the correction is of order $0.1b$, with $b$ being usually much larger than 1. But $\beta$ grows with $\gamma$, thus the convergence to the thermodynamic limit becomes faster, e.g., for $\gamma=4$ we have estimated $\beta>0.4$.

We investigated also the range $2<\gamma<3$. Again, we calculated moments for different values of $N$, for $\gamma=2.5$ and fitted the formula (\ref{expectedmu}). The theory \cite{ref:evans2} predicts $\alpha=1/(\gamma-1) = 2/3$ for this case. 
The agreement between the data and Eq.~(\ref{expectedmu}) is very good for $\beta\approx 0.33$.

At last, in order to check whether the formula (\ref{wxasympt}) for the large-$x$ behavior of the cutoff function holds for multigraphs, we fitted a function $A+\ln(1+Bx)-(x/C)^D$ to $\ln w(x)$ obtained from numerical integration for $N=100,200,51200$, for $\gamma=2.5,\,3,\,3.5$. The assumed form of $w(x)$ approximates the measured cutoff functions very well.
The $D$'s obtained for different sizes $N$ tend to some limiting values which we found to be $D_{N\to\infty}=2.0$ for $\gamma=2.5$, $1.9$ for $\gamma=3$ and $3.1$ for $\gamma=3.5$, with uncertainties of order $0.1$. These values are in good agreement with the exponent $\eta=1/(1-\alpha)$ from Eq.~(\ref{wxasympt}) which gives $2,2$, and $3$, respectively, and with theoretical results for the equivalent model \cite{ref:evans2}.

To summarize our findings for the degenerated graphs: for $\gamma=3$ we have observed logarithmic corrections to the scaling $k_{\rm cutoff}\sim N^\alpha$. 
The point $\gamma=3$ is the critical one; for $\gamma\neq 3$, corrections are power-like, with the exponent approaching zero for $\gamma\to 3$. 
The corrections are thus very strong for $\gamma\approx 3$, even for very large networks. We have also examined the large-$x$ behavior of the cutoff function $w(x)$ and showed that it agrees with Eq.~(\ref{wxasympt}) derived for growing networks. This suggests some universality, but we will see later that it does not hold for simple graphs.

\subsection{Equilibrated trees}
In Ref.~\cite{ref:bbjk}, the partition function $Z(N)$ for equilibrated trees is found to be
\bq
	Z(N) = \oint \frac{dz}{2\pi i} z^{-N-1} Z_{\rm GC}(z). \label{zfortrees}
\eq
Here $Z_{\rm GC}(z)$ is a grand-canonical partition function obeying the equation
\bq
	Z_{\rm GC}(z) = z \tilde{F}(Z_{\rm GC}(z)), \label{zgc}
\eq
with
\bq
	\tilde{F}(z) = \sum_{k=0}^\infty \pi_\infty(k+1)z^k , \label{Ftree}
\eq
assuming that in the thermodynamic limit we want to obtain the degree distribution $\pi_\infty(k)$.
With help of Eq.~(\ref{zgc}), the partition function (\ref{zfortrees}) can be rewritten as:
\bq
	Z(N) = \oint \frac{dz}{2\pi i} z^{-N}\tilde{F}^{N-1}(z)  \left(\tilde{F}(z)-z \tilde{F}'(z)\right).
\eq
Because we assume $\pi_\infty(0)=0$, the function $\tilde{F}(z)$ equals $\frac{1}{z}F(z)$, with $F(z)$ as in Eq.~(\ref{Fpseudo}) for multigraphs.
This in turn leads to the following formula for $Z(N)$:
\bq
	Z(N) = \frac{1}{N} \oint \frac{dz}{2\pi i} z^{-2N} F^N(z),
\eq
which is, up to a factor $1/N$, equivalent to the partition function (\ref{Zpseudo2}) for multigraphs with $\bar{k}=2-1/N$. 
This proves the equivalence between equilibrated multigraphs and trees, provided that the average degree is properly tuned, which has been already reported \cite{ref:bib}.
In our case, when $L=N$ for multigraphs and therefore $\bar{k}=2$, a small difference in the values of moments $\mu_m$ arise. This does not, however, change the fact that in the thermodynamic limit both models become fully equivalent, and that for any finite $N$ the difference can be neglected. 

This means that the degree distribution for equilibrated trees behaves exactly as for multigraphs, that is we again have logarithmic corrections to the moments $\mu_m(N)$ for $\gamma=3$, and power-law corrections for $\gamma\neq 3$.
This indicates also that the cutoff exponent $\alpha=1/2$ for $\gamma\geq 3$ and $\alpha=1/(\gamma-1)$ for $\gamma<3$. 

\section{Multicanonical simulations of simple graphs}
\label{sec:muca}

So far we have considered equilibrated multi- and tree graphs. But real-world networks are usually simple graphs, that is they do not have multiple- and self connections, and have loops. 
Unfortunately, for these reasons simple graphs are not accessible with the technique used before, because one does not know how to write the partition function as a single contour integral. Therefore, to obtain $\pi_N(k)$ one needs to turn to Monte Carlo (MC) techniques.
Let us shortly describe here the general method which serves for this purpose, details can be found elsewhere \cite{naszcpc,ref:homnasz}.
We simulate graphs with fixed number of nodes $N$ and links $L$. Each new graph $g_{t+1}$ is generated from the previous one  $g_t$ by  rewiring a single link. The move is accepted with the Metropolis probability:
\bq
	P(g_t\to g_{t+1}) = \min\left\{1,\frac{W(g_{t+1})}{W(g_t)} \right\}, \label{wbbj:metr}
\eq
which ensures that graphs are generated with correct weights $W(g)$ from Eq.~(\ref{wprod}).
In order to restrict to simple graphs we reject moves introducing self- or multiple connections.
This method allows for estimating $\pi(k)$ with good accuracy for $N$ of order thousands, which is, however, too small for our purpose. 
Larger networks are not accessible in a reasonable computer time. To show this, assume that $\pi(k)\sim k^{-\gamma}$ and that we start from a network with all $k$'s much smaller than $k_{\rm cutoff}\sim N^\alpha$. We assume that to obtain the experimental distribution $\pi_{\rm exp}(k)$ with acceptable level of noise, in the course of simulation each node has to change its degree between $1$ and $k_{\rm cutoff}$ many times. Let us denote by $T$ the first-passage time from $k=1$ to $k=k_{\rm cutoff}$. Our algorithm changes $k$ by $\pm 1$ every step, so the process can be treated as a biased random walk in a potential $V(k)=-\ln \pi(k)$. Then $T$ can be estimated as $T \sim N^{\alpha (\gamma+1)}$ for $\gamma>2$, which is nothing more than the Arrhenius law $T\sim \exp\left[V(k_{\rm cutoff})\right]$ with an additional correction factor. 

\begin{figure}
\includegraphics*[width=8cm]{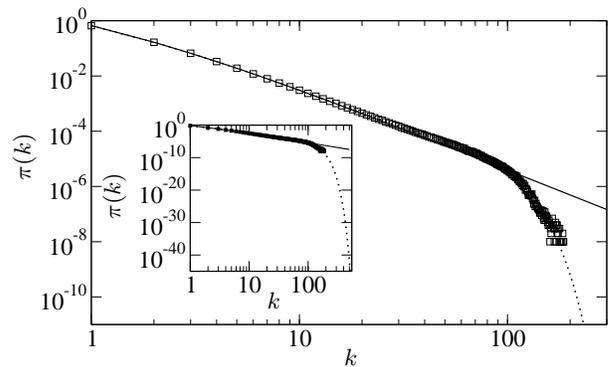}
\caption{\label{fig3}Example of $\pi(k)$ for simple MC (squares) and MUCA (dotted line) for simple graphs with $N=10^3$ and $\gamma=3$. Solid line shows $\pi_\infty(k)$. Inset: the same for a broader range of $k$. MUCA simulation allows to estimate the tail of $\pi_N(k)$ much better than simple MC.}
\end{figure}

If, however, one could ``flatten'' $\pi(k)$ by appropriate reweighting of graphs, then instead of a biased random walk one would deal with a simple random walk where $T\sim N^{2\alpha}$. This is always better than the previous estimate for the interesting range of $\gamma$.
Moreover, by increasing the probability of graphs with $k>k_{\rm cutoff}$, one could measure $\pi_{\rm exp}(k)$ far above the cutoff. 
Here comes the idea of multicanonical simulations (MUCA) \cite{ref:muca1}, a similar idea has already been applied to graphs \cite{ref:akh}. To apply MUCA in our case, we modify the weight function from Eq. (\ref{wprod}) to:
\bq
	W(g) = \frac{r(k_1)}{\pi_\infty(k_1)} \prod_{i=1}^N p(k_i), \label{wprodMUCA}
\eq
where $r(k_1)$ is some function depending on the degree of one node, say the first one, and is chosen so that the distribution $\pi_{\rm flat}(k)$, measured now only for the 1st node, is flat. The factor $1/\pi_\infty(k_1)$ gives additional advantage of flattening the distribution at no cost for $k\ll k_{\rm cutoff}$. To obtain the true distribution $\pi_{\rm exp}(k)$ one multiplies $\pi_{\rm flat}(k)$ by $\pi_\infty(k)/r(k)$. Since all nodes are statistically equivalent in equilibrated graphs, this procedure must give the same result as standard MC simulations. The only question is to find the optimal $r(k)$. In our simulations, we applied the iterative method from Ref.~\cite{ref:muca}, which calculates $r(k)$ ``on the fly'' from sampled data. Because changing $r(k)$ during the simulation violates  detailed balance and thus can change statistical weights, after obtaining sufficiently flat histogram of $\pi_{\rm flat}(k)$ we fix $r(k)$ and perform a simple MC run with $W(g)$ given by Eq.~(\ref{wprodMUCA}).
To speed up the simulation we divide the whole range of $k$, for which we want to determine $\pi(k)$, into overlapping subranges of size $\approx 50$ and then glue results. We validated the method for graphs by comparing it with direct MC simulations for $N\leq 1000$ (see Fig.~\ref{fig3}) and for multigraphs by comparing with exact results from Sec. III.A.

\begin{figure}
\includegraphics*[width=8cm]{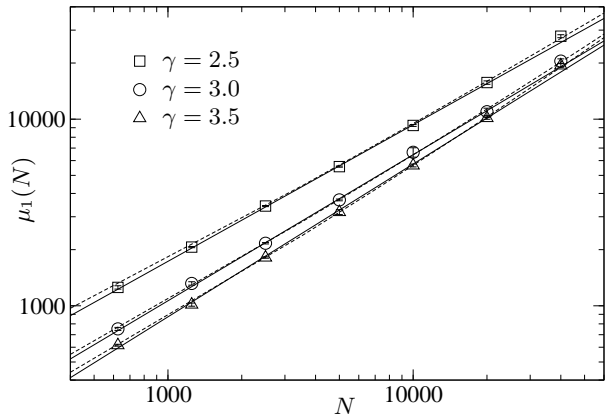}
\caption{\label{mu-graphs}Plots of $\mu_1(N)$ for simple graphs with $\gamma=2.5,3,3.5$. Solid and dotted lines are fits from Table \ref{tab1}, left and right, respectively.}
\end{figure}

To check how the cutoff scales with $N$, we simulated graphs with $N=625$, 1250, 2500, $\dots,40000$ nodes, and $\gamma=2.5,\, 3,\, 3.5$, for the same $\pi_\infty(k)$ as before. 
In each case we performed $10^{10}$ to $10^{11}$ MC steps (in units of single-link rewirings) for every subrange of $k$, after fixing $r(k)$. Each simulation was repeated 3 -- 4 times
in order to estimate statistical errors.
We then calculated the moments $\mu_m$ of the cutoff function and checked if one can fit a simple power law $\mu_m \cong aN^{\alpha(m+1)}$ to data points. Results are shown in Table \ref{tab1}, left, together with reduced $\chi^2$ values and significance levels (``$Q$'' values) for the fits. We see that none of these cases can be accepted with the confidence level $0.05$ which is commonly used in hypotheses testing. On the other hand, if we assume a more sophisticated form of $\mu_m(N)$ with the subleading correction from Eq.~(\ref{expectedmu}) and, according to Ref.~\cite{ref:dr3}, with $\alpha(\gamma\geq 3)=1/2, \; \alpha(\gamma=2.5)=1/(5-\gamma)=0.4$, the agreement is very good, see Table \ref{tab1}, right. 
We thus conclude that the subleading term plays an important role for moderate sizes.
In Fig.~\ref{mu-graphs} we show plots of $\mu_1(N)$ and both fitted functions.
The subleading correction coefficient $b$ is very large. For instance, for $\gamma=3.5$ the correction is of order $0.1$ for $N\sim 10^6$, much larger than for multigraphs, perhaps due to much stronger structural constraints. Therefore, simple graphs approach the thermodynamic limit even slower than equilibrated trees or degenerated graphs.

Finally, we examined the large-$x$ behavior of $w(x)$ using the same method as for multigraphs. We found that the exponent $\eta$ in Eq.~(\ref{wxasympt}) is not $1/(1-\alpha)$ as for multigraphs but approximately $2.0$ for all values of $\gamma$. This agrees with results of Ref.~\cite{ref:dr3} that the cutoff is always Gaussian in simple graphs, but it means also that $\eta$ is not universal. 

\begin{table}
\caption{\label{tab1}Results of fitting of the two different power laws (without and with a correction) to the measured moments $\mu_m$ for simple graphs. For each $\gamma$, three consecutive rows present results for $m=1,2,3$. ``Parameters'' indicate the set of free parameters during the fitting procedure, $Q$ is the significance level of the fit. If $\alpha$ is marked as ``fixed'', its value is $0.4$ for $\gamma=2.5$ and $0.5$ for the two other cases.}
\vspace{3mm}
\begin{tabular}{c|ccc|cccc}
\hline
\hline
		& \multicolumn{3}{c|}{$\mu_m=aN^{\alpha(m+1)}$} & \multicolumn{4}{c}{\hspace{2mm}$\mu_m=aN^{\alpha(m+1)}(1+bN^{-\beta})$\hspace{2mm}} \\
		& \multicolumn{3}{c|}{parameters $a,\alpha$} & \multicolumn{4}{c}{parameters $a,b,\beta$, fixed $\alpha$} \\
\hline
\hspace{2mm}$\gamma$\hspace{2mm} & \hspace{2mm}$\alpha$\hspace{2mm} & red. $\chi^2$ & $Q$ & \hspace{2mm}$\beta$\hspace{2mm} & \hspace{2mm}$b$\hspace{2mm} & red. $\chi^2$ & $Q$ \\
\hline
2.5	&		0.366   &		4.5		&	<0.001 		& 0.39	&	5.4	& 2.40	&	0.05	\\
		&		0.373		&		3.5		& 0.004 		& 0.42	& 8 & 1.99	& 0.09	\\
		&		0.376		&		2.9		&	0.013 		& 0.44	& 11	& 1.72	& 0.14 \\
\hline
3		&		0.392		&		2.64	&	0.021 		&	0.34	&	30	&	0.94	&	0.44 \\
		&		0.400		&		2.41	&	0.034 		& 0.40	&	80	&	0.87	&	0.48 \\
		&		0.405		&		2.34	&	0.039 		& 0.47	&	200	&	0.86	& 0.49 \\
\hline
3.5	&		0.409		&		6.1		& <0.001 		&	0.38	& 22	& 1.09	& 0.36 \\
		&		0.420		&		6.8		& <0.001 		& 0.48	& 56	& 0.73	& 0.57 \\
		&		0.428		&		9.2		&	<0.001 		& 0.56	&	125	& 0.51	& 0.72 \\
\hline
\hline
\end{tabular}
\end{table}

\begin{figure}
\includegraphics*[width=8cm]{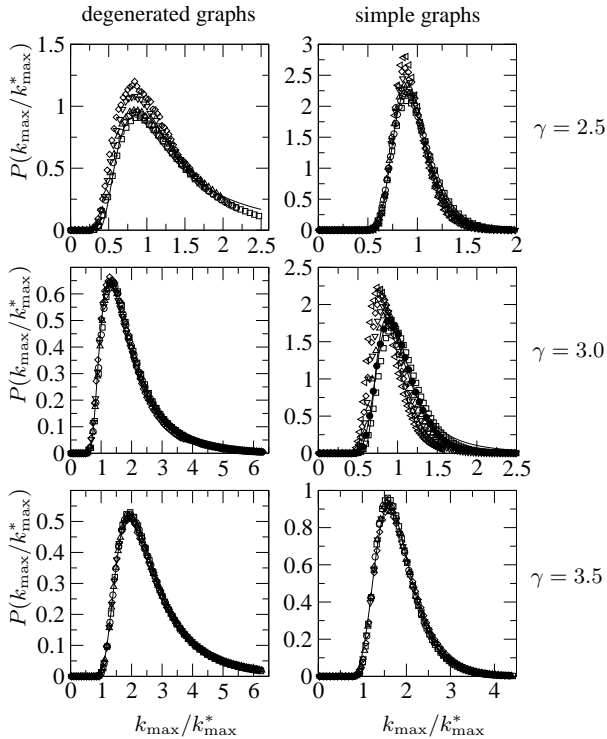}
\caption{\label{pkmax}Left: rescaled plots of $P(k_{\rm max})$ for degenerated graphs, for $\gamma=2.5$ (top), 3 (middle) and $3.5$ (bottom picture), and for different sizes ranging from $N=1000$ up to $32000$ (1000: square, 2000: circle, 4000: triangle up, 8000: triangle down, 16000: diamond, 32000: triangle left). Solid lines are Fr\'{e}chet distributions $P(x)\propto x^{-\gamma} e^{-x^{-\gamma+1}}$ fitted to data points for one chosen size $N$ (usually $N=4000$). Right: the same for simple graphs, lines are now: top and bottom -- Gumbel, middle -- Gumbel (thin line) and Fr\'{e}chet (thick line), fitted to black circles ($N=2000$).}
\end{figure}

\section{Cutoffs and the distribution of maximal degree}
So far we have studied the cutoffs in the degree distribution averaged over the ensemble of random graphs of a certain kind.
The value $k_{\rm cutoff}\sim N^\alpha$ tells where the power law ends and $\pi_N(k)$ starts to fall off rapidly. 
One can also pose the question: how does the maximal degree, $k_{\rm max}$, scale with $N$? 
Let $\alpha$ and $\alpha'$ be exponents in the power laws $k_{\rm cutoff}\sim N^\alpha$ and $k_{\rm max}\sim N^{\alpha'}$, respectively. The exponents $\alpha, \alpha'$ may, but do not have to be equal, which has not been emphasized strongly enough in the literature on complex networks.
To show this, let us consider the hypothetic network where all degrees $k_1,\dots,k_N$ are independent random variables taken from the distribution $p(k)\sim k^{-\gamma}$.
The degree distribution $\pi(k)$ is then equal to $p(k)$. In a real, finite network, degrees are (at least slightly) correlated due to structural constraints and $\pi(k)$ has a cutoff. In our simplified model, we can mimic these constraints by introducing the cutoff explicitly: $p(k)\sim k^{-\gamma} w(\frac{k}{N^\alpha})$ with some exponent $\alpha$. Let us focus on a particular form of the cutoff:
\bq
	w(x)=\exp(-a x^\eta),
\eq
with parameters $a,\eta$. This form of the cutoff function seems to be very general: in case of growing networks it is an asymptotic behavior for large $k$, we also checked numerically that it holds for equilibrated graphs. 
For large $N$, the distribution $P(k_{\rm max})$ of the maximal degree,
\bq
	P(k_{\rm max}) \approx N p(k_{\rm max}) \left( \int_{1}^{k_{\rm max}} p(k) dk \right)^{N-1}, \label{pkmaxgen}
\eq
takes its maximal value at $k^*_{\rm max}$ being a solution to
\bq
	p'(k^*_{\rm max}) \int_{1}^{\infty} p(k) dk  \cong -N p^2(k^*_{\rm max}),
\eq
which for the assumed cutoff gives
\bq
	k^*_{\rm max} \sim \left\{ \begin{tabular}{lc} $N^{\frac{1}{\gamma-1}}$, & $\alpha\geq\frac{1}{\gamma-1}$, \\
																							 $N^\alpha (\ln N)^{1/\eta}$, & $\alpha<\frac{1}{\gamma-1}$. \end{tabular} \right.
																							 \label{kmaxgen}
\eq
This means that, modulo the logarithmic correction, $\alpha'=\min(\alpha,1/(\gamma-1))$.
One can also show that $P(k_{\rm max})$ for $\alpha>1/(\gamma-1)$ is given by the Fr\'{e}chet extreme value distribution: $P(x)\propto x^{-\gamma} e^{-x^{-\gamma+1}}$, while it approaches the Gumbel distribution: $P(x) \propto e^{-x-e^{-x}}$ for $\alpha<1/(\gamma-1)$, in the properly rescaled variable $x=A+B k_{\rm max}$. 
Because these two distributions are narrow, $k^*_{\rm max}$ approximates also the mean value $\left<k_{\rm max}\right>$.
We can learn two things from Eq.~(\ref{kmaxgen}). 
First, $k_{\rm max}$ scales as a pure power of $N$ with the same exponent as for $k_{\rm cutoff}$, $\alpha' = \alpha$, only if $\alpha=1/(\gamma-1)$.
The distribution of $k_{\rm max}$ is then neither Fr\'echet nor Gumbel, but has a more complicated form, as follows from Eq.~(\ref{pkmaxgen}):
\bq
	P(x) \cong \mathcal{N} x^{-\gamma} w(x) \exp\left[ -\mathcal{N} \int_x^\infty y^{-\gamma} w(y) dy \right]
\eq	
where $\mathcal{N}$ is a normalization coefficient and $x=k_{\rm max}/N^{1/(\gamma-1)}$, and depends on the exact form of $w(x)$.
Second, $\alpha'$ may be smaller than $\alpha$, that is the maximal degree grows slower than the cutoff.
At first sight this may appear counter-intuitive: one could have expected that $\pi(k)$ for large $k$ is dominated by the maximal degree distribution and thus $k_{\rm max}$ must not grow slower than $k_{\rm cutoff}$.

The formula for $k^*_{\rm max}$ presented above 
works surprisingly well for real graphs where the cutoff stems from correlations between nodes degrees. In order to check Eq.~(\ref{kmaxgen}), we performed multicanonical simulations and obtained distributions $P(k_{\rm max})$ for multigraphs and simple graphs with $\gamma=2.5, \; 3$ and $3.5$.
In light of what has been said in Sec. III.B, the scaling of $k_{\rm max}$ for trees has to be identical to that for multigraphs, so it is not necessary to consider trees separately. The procedure was similar to that described in the previous section, with the only difference that we flatten $P(k_{\rm max})$ and not $\pi(k)$. In each simulation we performed around $10^8$ Monte Carlo steps after fixing the weight $r(k_{\rm max})$.

Let us discuss first multigraphs. In Table \ref{tab2}, left, we present values of $\alpha,\eta$ for various $\gamma$, which we confirmed numerically in previous sections. The fourth column shows scaling laws for $k^*_{\rm max}$ predicted by means of Eq.~(\ref{kmaxgen}).
In Fig.~\ref{pkmax} we plotted the experimental distributions $P(k_{\rm max})$, in the rescaled variable $x=k_{\rm max}/k^*_{\rm max}$. Plots for different $N$ show good agreement of positions of the maximum. The distribution for $\gamma=3.5$ is also approximately Fr\'{e}chet, which agrees with recent findings \cite{ref:majumdar}, the two for $\gamma=2.5, \; 3$ deviate slightly from Fr\'{e}chet in the tail.
The perfect scaling of $k_{\rm max}$ means that the thermodynamic limit for the distribution of maximal degree is approached much faster than for $\pi(k)$, and is reached already for $N\sim 10^4$ nodes; no subleading corrections are necessary.

We repeated the same procedure for graphs, see Table \ref{tab2}, right, and Fig.~\ref{pkmax}. Again, one sees very good agreement, except for $\gamma=3$, where one obtains better collapse for a slightly different value $k^*_{\rm max}\sim N^{0.45}$. The shape of $P(k_{\rm max})$ is better approximated by Gumbel than Fr\'{e}chet distribution. 
Also for $\gamma=3.5$ and $N\leq 32000$, the distribution $P(k_{\rm max})$ deviates from Fr\'{e}chet and is closer to Gumbel.

\begin{table}
	\caption{\label{tab2}Values of exponents $\alpha,\eta$, and scaling of $k^*_{\rm max}$ for degenerated and simple graphs, for different power laws $\pi(k)\sim k^{-\gamma}$. Assumed scalings are: $k_{\rm cutoff}\sim N^\alpha$, $\ln w(x)\sim -x^\eta$. The exponents $\alpha,\eta$ have been confirmed numerically in the previous section. The formulas for $k^*_{\rm max}$ are obtained from Eq.~(\ref{kmaxgen}).}
	\vspace{3mm}
	\begin{tabular}{c|ccc|ccc}
		\hline
		\hline
		\hspace{2mm} $\gamma$	\hspace{2mm} & \multicolumn{3}{c|}{multigraphs} & \multicolumn{3}{c}{simple graphs} \\
		\hline
				& \hspace{2mm} $\alpha$ \hspace{2mm}	& \hspace{2mm}$\eta$\hspace{2mm} 	&	$k^*_{\rm max}$ 	&	\hspace{2mm} $\alpha$ \hspace{0mm}	& $\hspace{0mm}\eta$\hspace{0mm} 	&	$k^*_{\rm max}$ \\
		\hline
		$(2,3)$	&	$\frac{1}{\gamma-1}$		&	$\frac{\gamma-1}{\gamma-2}$			&	$N^{1/(\gamma-1)}$		&	$\frac{1}{5-\gamma}$		&	2				&	$N^{1/(5-\gamma)} (\ln N)^{1/2}$	\\
		$3$		&	1/2+log	&	2				&	$N^{1/2}$		&	1/2				& 2				&	$N^{1/2}$								\\
		$(3,\infty)$	&	1/2				&	2				&	$N^{1/(\gamma-1)}$		&	1/2				&	2				&	$N^{1/(\gamma-1)}$								\\
		\hline
		\hline
	\end{tabular}
\end{table}

To conclude, we have shown that the distribution of the maximal degree exhibits the scaling $k_{\rm max} \sim N^{\alpha'}$ but in general with a different exponent than that in the cutoff. 
The large-$N$ scaling is, however, approached much faster than in case of the degree distribution and subleading corrections can be safely neglected in most cases for networks of order $10^4$ nodes.
It is also worth mentioning that $\alpha'$ depends on both the exponents $\gamma$ and $\alpha$.

\section{Summary}
In this paper we considered finite-size effects in the degree distribution of equilibrated networks.
We showed that the convergence towards the thermodynamic limit is very slow, thus in order to get reasonable estimation of the cutoff one has to consider subleading corrections to the scaling $k_{\rm cutoff}\sim N^\alpha$. For multigraphs and $\gamma=3$, the correction turns
into a leading behavior $\sim (N\ln N)^{1/2}$ and therefore does not vanish for $N\to\infty$. This has to be taken into account when comparing any numerical results for finite networks to that derived analytically for the leading behavior only.
We checked also that the asymptotic behavior of the cutoff function is very simple: $\ln w(x) \sim -x^\eta$, but $\eta$ is not universal among different classes of graphs.
We argued that the maximal degree $k_{\rm max}$ reaches the asymptotic power-law scaling faster than the cutoff but scales in many cases differently than $k_{\rm cutoff}$.
 
The networks presented here do not have explicit degree-degree correlations. It would be interesting to check which of the results will survive the introduction of correlations, i.e., are universal to some extent.

\section*{Acknowledgments}
B. W. thanks A. Krzywicki and S. N. Majumdar for discussions. We are particularly indebted to Z. Burda for many useful comments. We also thank the EC-RTN Network “ENRAGE” under grant No.~MRTN-CT-2004-005616 and the Alexander von Humboldt Foundation for support.

\end{document}